\documentclass[a4paper]{jpconf}

\usepackage[english]{babel} 
\usepackage[latin1]{inputenc}
\usepackage{color}
\usepackage{graphicx}
\usepackage{amsmath,amssymb,amsfonts}
\usepackage[margin=2.5cm]{geometry}
\usepackage[colorlinks=true,linkcolor=blue,urlcolor=blue,citecolor=blue]{hyperref}
\usepackage[percent]{overpic}

\newcommand {\mc} {\mathcal}
\newcommand {\lan} {\left \langle}
\newcommand {\ran} {\right \rangle}

\newcommand {\sign} {\mathrm{sign}}

\begin{document}
	
\title{Photoassisted shot noise spectroscopy at fractional filling factor}

\author{Luca Vannucci$^{1,2}$, Flavio Ronetti$^{1,2,3}$, Dario Ferraro$^3$, J\'er\^ome Rech$^3$, Thibaut Jonckheere$^3$, Thierry Martin$^3$ and Maura Sassetti$^{1,2}$}

\address{$^1$Dipartimento di Fisica, Universit\`a di Genova, Via Dodecaneso 33, 16146, Genova, Italy}
\address{$^2$CNR-SPIN, Via Dodecaneso 33, 16146, Genova, Italy}
\address{$^3$Aix Marseille Univ, Université de Toulon, CNRS, CPT, Marseille, France}

\ead{vannucci@fisica.unige.it}

\begin{abstract}

We study the photoassisted shot noise generated by a periodic voltage in the fractional quantum Hall regime. Fluctuations of the current are due to the presence of a quantum point contact operating in the weak backscattering regime. We show how to reconstruct the photoassisted absorption and emission probabilities by varying independently the dc and ac contributions to the voltage drive. This is made possible by the peculiar power-law behavior of the tunneling rates in the chiral Luttinger liquid theory, which allow to approximate the typical infinite sums of the photoassisted transport formalism in a simple and particularly convenient way.

\end{abstract}

\section{Introduction}

With the ultimate goal of controlling coherent few-particle excitations in quantum conductors, the condensed matter community has been paying an ever increasing attention to ac transport in mesoscopic devices over the last twenty years. This has lead to the development of full counting statistics \cite{Levitov93,levitov96}, photoassisted transport formalism \cite{Pedersen98,Platero04}, Floquet scattering matrix approach \cite{Moskalets02} and challenging high-frequency experimental techniques \cite{Reydellet03,Feve07,Gabelli13,dubois13-nature}, culminated in the new paradigm of electron quantum optics (EQO) \cite{Bocquillon12,Bocquillon14}.

However, less attention has been devoted to the role of electron-electron interactions, which often play a crucial role in low-dimensional systems. For instance, the fractional quantum Hall (FQH) effect emerges as a consequence of strong repulsive interactions that give rise to exotic quasi-particles carrying fractional charge and statistics \cite{Laughlin83}. At the same time, the presence of dissipationless topological edge modes makes the FQH phase a good candidate to study photoassisted quantum transport in interacting systems \cite{crepieux04,Dolcetto14,Rech16,Vannucci17}.

In this paper we investigate the photoassisted shot noise (PASN) in the FQH regime. We consider a FQH system where periodic voltage pulses are injected from one of the terminals in presence of a quantum point contact (QPC). In this geometry, excitations incoming from the leads are partitioned at the QPC, in a protocol that is reminiscent of the Hanbury-Brown and Twiss optical experiment \cite{Bocquillon12}. We evaluate the shot-noise at the first relevant order in the tunneling, considering the QPC in a weak backscattering regime. 
While the expression for the shot noise generally consists of an infinite superposition of dc contributions, each one weighted by the corresponding photoassisted probability, we show that the FQH physics allows to extract each single contribution to the PASN in a surprisingly simple fashion. We provide a recipe to reconstruct the typical absorption and emission probabilities of the photoassisted formalism by independently tuning the ac and dc components of the voltage drive. We also discuss the experimental feasibility of this study, identifying a set of reasonable experimental parameters under which our protocol should be applicable.

\section{Model}

We consider the edge states of a quantum Hall system with filling factor in the Laughlin sequence $\nu=1/(2n+1)$, $n \in \mathbb N$ \cite{Laughlin83}. The low energy excitations of such a quantum state are described by the chiral Luttinger liquid theory \cite{Wen90}, with a pair of counter-propagating chiral modes (one for each edge of the sample) represented by bosonic fields $\phi_{R/L}(x)$. For an infinite Hall bar they satisfy $[\phi_{R/L}(x), \phi_{R/L}(y)] = \pm i \pi \sign(x-y)$. They are linked to creation and annihilation of fractional quasi-particles with charge $e^* = \nu e$ through the bosonization identity
\begin{equation}
	\label{eq:bosonization}
	\psi_{R/L}(x) = \frac{U_{R/L}}{\sqrt{2\pi a}} e^{\pm i k_{\rm F} x} e^{-i \sqrt \nu \phi_{R/L}(x)} ,
\end{equation}
where the parameter $a$ is a short-distance cutoff and $U_{R/L}$ are the Klein factors. The quasi-particle density operators read $\rho_{R/L} (x) = \mp \frac{\sqrt{\nu}}{2\pi} \partial_x \phi_{R/L}(x)$ and the effective low-energy Hamiltonian is given by ($\hbar=1$)
\begin{equation}
	H = \frac{\pi v}{\nu} \int_{-\infty}^{+\infty} dx \left[ \rho^2_R(x) + \rho^2_L(x) + \Theta(-x-d) V(t) e \rho_R(x) \right]  ,
\end{equation}
with $v$ the propagation velocity of the free chiral fields. Here, we have included an additional coupling between the right-moving density and the time dependent voltage drive $V(t)$, which is used to drive the system out of equilibrium. The voltage $V(t)$ is assumed to be spatially homogeneous in the interval $(-\infty,-d)$, with $d>0$. It is easy to verify that the bosonic field $\phi_R(x)$ in the presence of $V(t)$ acquires an additional term, namely
\begin{equation}
\label{eq:sol_eq_motion}
	\phi_R(x,t) = \phi_R^{(0)}(x-vt,0) + e \sqrt \nu \int_0^{t - \frac x v - \frac d v} dt' V(t') ,
\end{equation}
where $\phi_R^{(0)}(x,t)$ is the field at equilibrium [i.e.\ $V(t)=0$]. To model tunneling between the upper and the lower edge we introduce a QPC at $x=0$, represented by the bosonized tunneling Hamiltonian
\begin{equation}
\label{eq:H_tun}
	H_{\rm tun}(t) = \lambda \exp \left[ i e^* \int_0^t dt' \, V(t') \right] e^{i \sqrt \nu \phi_R^{(0)}(0,t)} e^{-i \sqrt \nu \phi_L^{(0)}(0,t)} + \mathrm{h.c.}
\end{equation}
The constant $\lambda$ plays the role of the bare tunneling amplitude [in which we have also absorbed the factor $1/(2\pi a)$], while the time dependent phase is generated by the voltage drive as shown in Eq.\ \eqref{eq:sol_eq_motion}. It is worth noticing that we have neglected the delay $d/v$ due to the finite distance between the contact and the QPC, which has no effect on the time-averaged quantities we are interested in. We have also omitted the unnecessary Klein factors \cite{Guyon02}.

We will consider two types of periodic voltage drive, with period $T=2\pi/\omega$. The first one is the sinusoidal wave $V_{\rm sin}(t) = V_{\rm dc} - V_{\rm ac} \cos(\omega t)$ while the second one is a Lorentzian drive given by
\begin{equation}
	V_{\rm Lor} = V_{\rm dc} + V_{\rm ac} \left[ \frac 1 \pi \sum_{k=-\infty}^{+\infty} \frac{\eta}{\eta^2 + (t/T+k)^2}- 1 \right] .
\end{equation}
The parameter $\eta$ gives the half-width of each pulse, $W=\eta T$. The Lorentzian voltage represents a particularly interesting choice since it gives rise to minimal excitation states both in the integer and the FQH regime, called levitons \cite{dubois13-nature,Rech16,keeling06}.
Following a common procedure in the framework of the photoassisted transport \cite{Platero04,Tien63}, we write the exponential in Eq.\ \eqref{eq:H_tun} as a Fourier series
\begin{equation}
\label{eq:series}
	\exp \left\{ -i e^* \int_0^t dt' \, V(t') \right\} = e^{-i q \omega t}\sum_{l=-\infty}^{+\infty} p_l(\alpha) e^{-i l \omega t} ,
\end{equation}
where $q=e^* V_{\rm dc}/\omega$ and $\alpha=e^* V_{\rm ac}/\omega$ are linked to the dc and ac amplitudes respectively.
Focusing on the signals introduced above we get $p_l(\alpha)=J_{-l}(\alpha)$ \cite{crepieux04} for the sinusoidal voltage [$J_{n}(z)$ is the Bessel function of the first kind] and
\begin{equation}
	p_l(\alpha) = \alpha e^{-2\pi l \eta} \sum_{s=0}^\infty (-1)^{s} \frac{\Gamma(l+s+\alpha)}{\Gamma(1-s+\alpha)} \frac{e^{-4\pi s \eta}}{s! (s+l)!}
\end{equation}
for the Lorentzian drive \cite{dubois13,grenier13} .

\section{Backscattering current and noise}

In the weak backscattering regime, the QPC is almost fully transparent and only a small portion of the current is reflected back. Such a backscattering current can be calculated perturbatively at the first relevant order in the tunneling. We define the quasi-particle current operator as $I_B=e^* \dot N_L$, with the number of quasi-particle in the lower edge given by $N_L = \int dx \rho_L(x)$. This gives
\begin{equation}
	I_B(t) = i e^* \lambda \exp \left[ i e^* \int_0^t dt' \, V(t') \right] e^{i \sqrt \nu \phi_R^{(0)}(0,t)} e^{-i \sqrt \nu \phi_L^{(0)}(0,t)} + \mathrm{h.c.} 
\end{equation}
Fluctuations of the backscattering current are encoded in the zero-frequency shot noise defined as
\begin{equation}
	\mc S = 2\int_0^T \frac{dt}{T} \int_{-\infty}^{+\infty} dt' \left[ \lan I_B(t) I_B(t') \ran - \lan I_B(t) \ran \lan I_B(t') \ran \right] .
\end{equation}
To calculate $\lan I_B \ran$ and $\mc S$ one is asked to evaluate quantum averages involving quasi-particle fields. Invoking the bosonization identity Eq.\ \eqref{eq:bosonization}, they can be related to the bosonic Green's function $\mc G(\tau)$, which is equal for left and right moving particles and is given by ($\kappa_{\rm B}=1$) \cite{Ferraro10_PRB,Ferraro10_NJP}
\begin{equation}
	\mc G(\tau) = \lan \phi(\tau) \phi(0)\ran - \lan \phi^2(0)\ran = \ln \left[ \frac{\pi \theta \tau }{\sinh \left( \pi \theta \tau \right) (1+i \omega_{\rm c} \tau)} \right] .
\end{equation}
Here we have introduced the high-frequency cutoff $\omega_{\rm c} = v/a$ and the temperature $\theta$, and we consider the limit $\theta/ \omega_{\rm c} \ll 1$.
Finally, we make use of the Fourier transform
\begin{equation}
	\hat P_g(E) = \int_{-\infty}^{+\infty} dt \, e^{i E t} e^{g \mc G(t)} = \left(\frac{2\pi \theta}{\omega_{\rm c}}\right)^{g-1} \frac{e^{E/(2\theta)}}{\Gamma(g) \omega_{\rm c}} \left|\Gamma\left(\frac{g}{2}-i \frac{E}{2\pi \theta}\right)\right|^2
\end{equation}
and the series in Eq.\ \eqref{eq:series} to get
\begin{equation}
\label{eq:S_photoassisted}
	\mc S = 2(e^*)^2 |\lambda|^2 \sum_{l=-\infty}^{+\infty} \left|p_l(\alpha)\right|^2 \left\{ \hat P_{2\nu} \left[ \left(q + l \right) \omega \right] + \hat P_{2\nu} \left[ - \left(q + l \right) \omega \right]\right\} .
\end{equation}
This is the photoassisted expression for the shot noise. It can be viewed as a superposition of several dc contributions, whose effective bias is shifted by an amount $l\omega$ with respect to the dc value $q\omega$ and weighted by a probability $|p_l(\alpha)|^2$, which is nothing but the probability for a quasi-particle to absorb or emit $l$ energy quanta \cite{dubois13}. We note that the ac and dc amplitudes are well separated in Eq.\ \eqref{eq:S_photoassisted}. Indeed, the former emerges as the argument $\alpha$ of the coefficients $p_l$, while the latter appears in the functions $\hat P_{2\nu}(E)$ via the parameter $q$.

\section{Results}

In what follows we study the behavior of the PASN given in Eq.\ \eqref{eq:S_photoassisted} when $q$ and $\alpha$ are varied independently. To begin, let us first discuss a set of reasonable values for $\theta$, $\omega$ and $\eta$. Experiments testing levitons in two-dimensional electron gases are usually performed at $\theta=10 - 100$ mK \cite{dubois13-nature,Jullien14,Glattli16_physE}, which also happens to be a range of temperature where well defined FQH states can be spotted \cite{Tsui99}. The Lorentzian voltage drive usually operates at a frequency $f=\frac{\omega}{2\pi}=5-6$ GHz, with dimensionless width of each pulse $\eta=0.1-0.2$ \cite{dubois13-nature,Jullien14,Glattli16_physE}. Higher frequencies are also used for the sinusoidal wave \cite{dubois13-nature}, and photoassisted transport in graphene nanoribbons illuminated with THz radiation was recently reported  \cite{Parmentier16}. We initially set $\eta=0.1$ and $\theta=0.1 \, \omega$, but lower values of the ratio $\theta/\omega$ can be reached in principle and will be discussed later.

\begin{figure}
	\centering
	\includegraphics[width=0.9\linewidth]{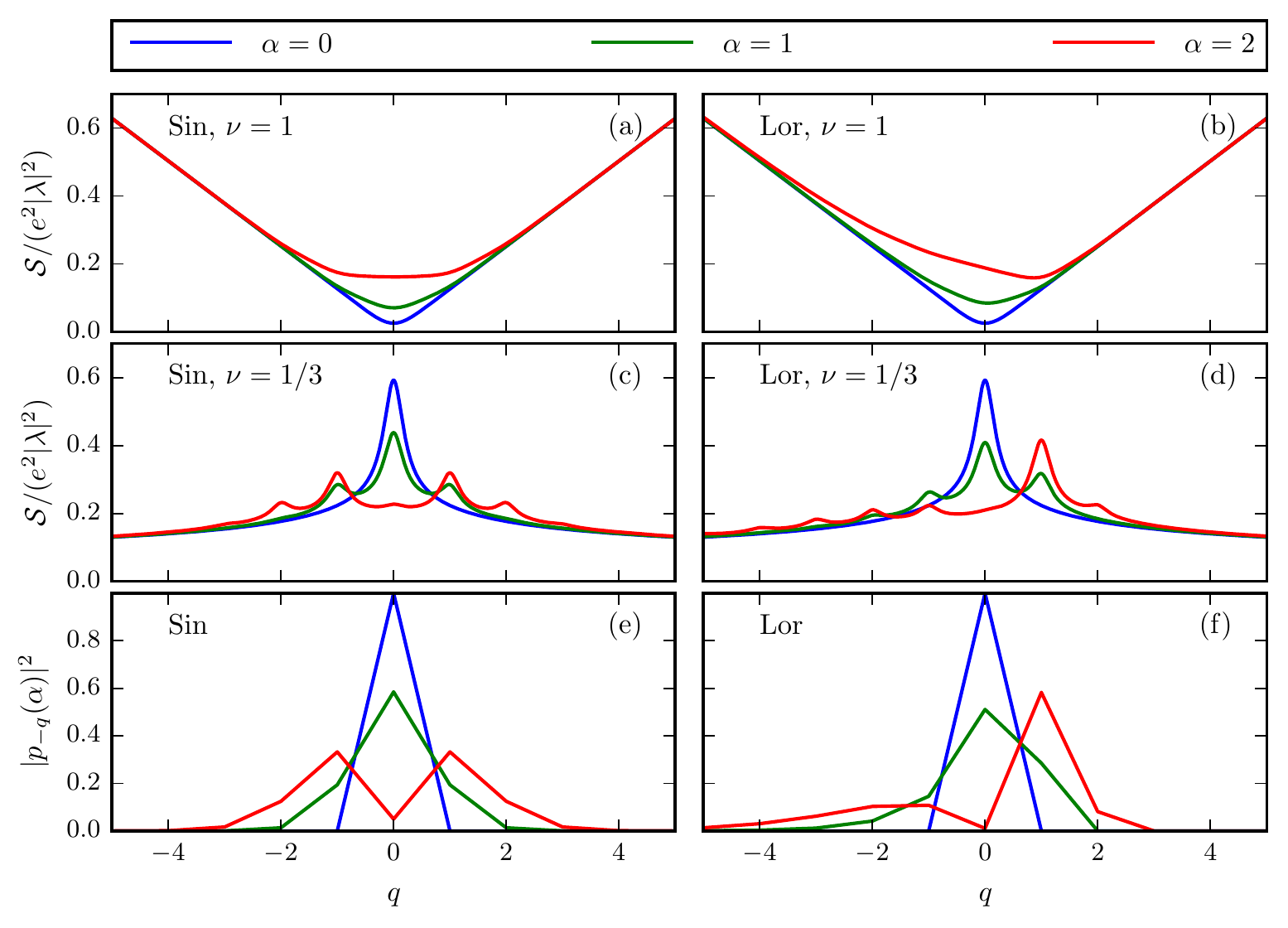}
	\caption{(a)-(d) PASN as a function of the dc amplitude $q$ for $\nu=1$ and $\nu=1/3$ and different values of the ac amplitude $\alpha$. (e)-(f) Coefficients $|p_{-q}(\alpha)|^2$ for different values of $\alpha$ as a function of $q$. In (e) and (f) the continuous lines are guides for the eyes, with the coefficients  $|p_{-q}(\alpha)|^2$ only defined for integer values of $q$.}
	\label{fig:fixed_ac}
\end{figure}

Figure \ref{fig:fixed_ac} shows the behavior of $\mc S$ as a function of $q$ for fixed values of $\alpha$. In panels (a) and (b) we report the case of sinusoidal and Lorentzian voltage pulses at integer filling factor $\nu=1$. The blue curve represents the pure dc case where no ac component is present in the voltage drive ($\alpha=0$). In this case $\mc S$ grows linearly with $q$ as expected, since $\mc S \propto q \coth(\frac{q\omega}{2\theta}) \approx |q|$ in the integer quantum Hall regime at sufficiently low temperature. Conversely, when a finite ac component is present ($\alpha=1,2$) the behavior at low $q$ is clearly non-linear and some excess noise due to the presence of the oscillating drive can be identified. Switching to the FQH regime [panels (c) and (d)], the linear (or almost linear) profile of the integer case is replaced by a strongly non-linear behavior, even for a dc voltage drive. This is a typical signature of the chiral Luttinger liquid theory. In particular the $\alpha=0$ curve, which is proportional to $\hat P_{2\nu}(q \omega) + \hat P_{2\nu}(-q \omega)$, shows a sharp peak around $q=0$. Such a structure is visible for $\alpha=1$ and $\alpha=2$ as well, with additional peaks arising for integer values of $q$. The different peaks in Fig.\ \ref{fig:fixed_ac} (c) and (d) reproduce the features of the $\alpha=0$ curve at shifted values $q+l$, since the photoassisted transport can be interpreted as an infinite superposition of shifted dc cases weighted by the probabilities $|p_l(\alpha)|^2$, as we mentioned before. Indeed, due to the sharply peaked structure of the tunneling rates in the FQH regime, the dominant contribution to the noise around integer $q$ is given by the photoassisted amplitudes with $l=-q$. In such a case the PASN is well approximated by
\begin{equation}
\label{eq:approx}
	\mc S \approx 4(e^*)^2 |\lambda|^2 \hat P_{2\nu} (0) \left|p_{-q}(\alpha)\right|^2 ,
\end{equation} 
allowing to reconstruct the probabilities $|p_l(\alpha)|^2$ from the relative height of the different peaks. Thus, by fixing the ac component of the voltage drive and tuning the dc component we can explore all the coefficients $|p_l(\alpha)|^2$ for $l=0,\pm 1, \pm 2 \dots$.
This is similar to the spectroscopic protocol developed by Dubois \emph{et al.}\ in Refs.\ \cite{dubois13-nature,dubois13} for the free Fermion case, although the fractional regime treated in the present work makes it much more effective and easy to visualize, due to the peculiar structure of the tunneling rate $\hat P_{2\nu}(E)$ at fractional filling (see also a similar analysis in the framework of finite frequency noise spectroscopy discussed in Refs.\ \cite{Carrega12,Ferraro14_NJP}).
For instance, Fig.\ \ref{fig:fixed_ac} (d) suggests that all $p_{-q}(\alpha)$ with $q>\alpha$ vanish for a Lorentzian drive with integer $\alpha$, since we cannot see any further peak at $q>\alpha$. This is the striking property that allows the Lorentzian voltage pulse to generate a single electron above the Fermi level, with no disturbance below it \cite{levitov96,keeling06}. Conversely, no cancellation arises in the sinusoidal case, where Fourier coefficients manifestly satisfy $|p_{q}(\alpha)|^2=|p_{-q}(\alpha)|^2$ [see Fig.\ \ref{fig:fixed_ac} (c)].
To check the validity of our spectroscopic protocol we also show the coefficients $|p_{-q}(\alpha)|^2$ for $\alpha=0,1,2$ in Fig.\ \ref{fig:fixed_ac} (e) and (f). One can easily see that the relative heights of all the peaks in Fig.\ \ref{fig:fixed_ac} (c) and (d) are very well reproduced by the coefficients $|p_{-q}(\alpha)|^2$. As an example, the absence (almost total) of peaks at $q=0$ for both the sinusoidal and the Lorentzian drive with $\alpha=2$ is linked to the fact that $|p_0(2)|^2 \ll 1$ for both signals. Moreover, the value of $|p_{-1}(2)|^2$ explains the high asymmetric peak at $q=1$ for the Lorentzian drive with $\alpha=2$. 

\begin{figure}
	\centering
	\includegraphics[width=0.9\linewidth]{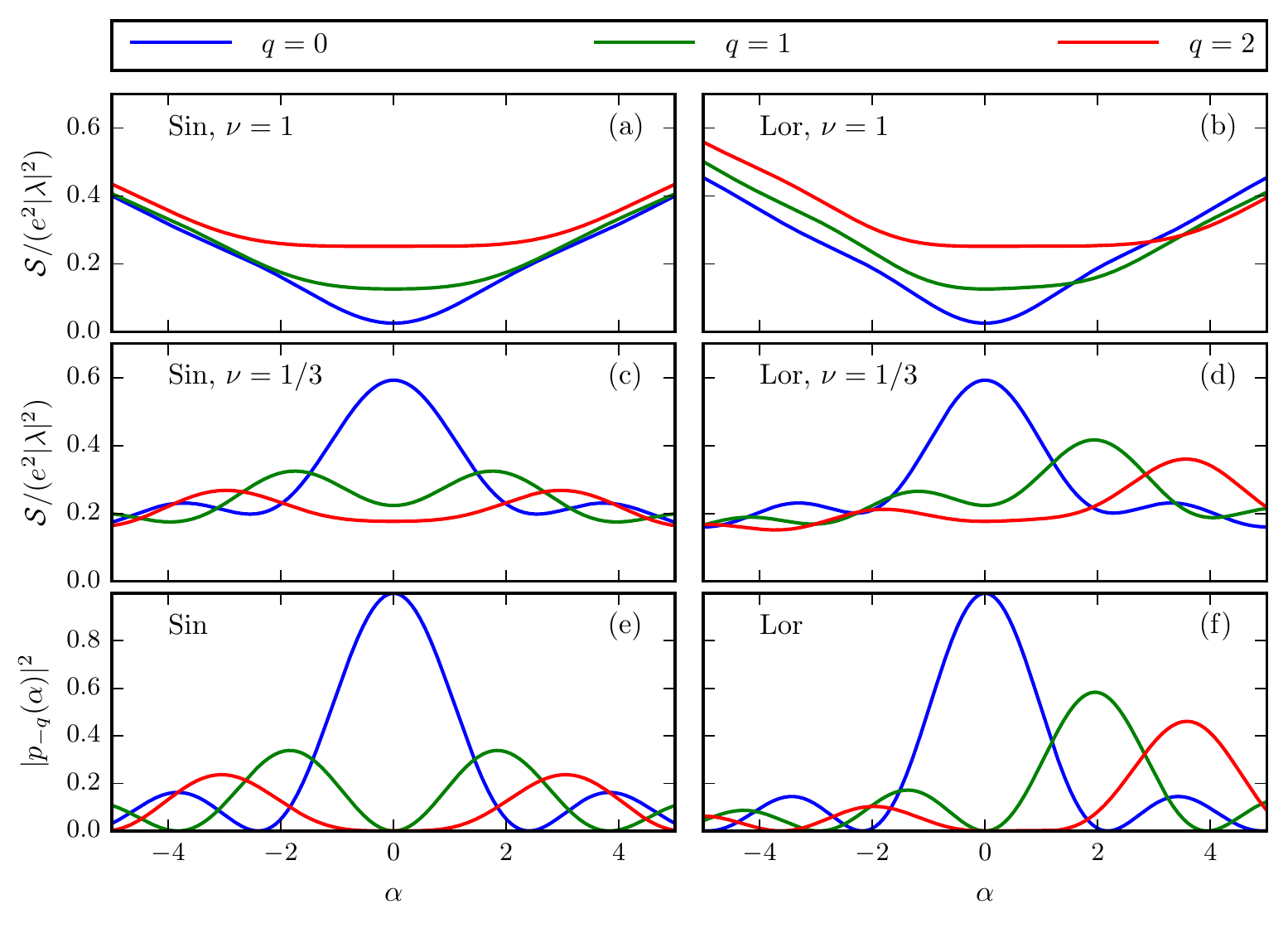}
	\caption{(a)-(d) PASN for $q=0,1,2$ as a function of $\alpha$ at $\nu=1$ and $\nu=1/3$. (e)-(f) Coefficients $|p_{-q}(\alpha)|^2$ for $q=0,1,2$ as a function of $\alpha$.}
	\label{fig:fixed_dc}
\end{figure}

We now turn to the opposite case, in which the dc component is fixed and we allow the parameter $\alpha$ to vary continuously. As shown in Fig.\ \ref{fig:fixed_dc} (a) and (b), at $\nu=1$ we get a linear behavior at high values of $|\alpha|$ both for the sinusoidal and the Lorentzian drive. In the vicinity of $\alpha=0$, the curves deviate from the linear regime and the noise is more or less proportional to $q$. We note once again the sharp asymmetry for $q \ne 0$ of the Lorentzian voltage drive, as opposed to the symmetric profile of the sinusoidal wave.
For fractional filling factor $\nu=1/3$ [panels (c) and (d)] the curves are evidently non-linear and oscillate in a non-monotonous fashion as a function of $\alpha$.
However, the behavior at $\nu=1/3$ is much more interesting since we can link the value of $\mc S$ to the probability $\left|p_{-q}(\alpha)\right|^2$, following Eq.\ \eqref{eq:approx}. 
In contrast with the case of fixed ac component, in this case we can explore the dependence of the $q$-th Fourier coefficients upon its argument $\alpha$. Comparing with Fig.\ \ref{fig:fixed_dc} (e) and (f), where we report the coefficients $|p_{-q}(\alpha)|^2$ as a function of $\alpha$, we observe that the approximation works well from a qualitative point of view, although an additional contribution due to finite temperature effects is present in all curves at $\nu=1/3$, thus preventing us from getting a good quantitative match.

In order to improve the spectroscopic protocol from a quantitative point of view, let us remark that the peculiar peak of the function $\hat P_{2\nu}(E) + \hat P_{2\nu}(-E)$ around $E=0$ becomes more and more pronounced as the ratio $\theta/\omega$ decreases. Indeed, it is well known that the Luttinger liquid theory can lead to a diverging power-law behavior in the limit $\theta \to 0$. Thus, the approximation Eq.\ \eqref{eq:approx} becomes more efficient at lower temperatures (or higher frequencies), since the relative weight of the term $l=-q$ in the PASN with respect to all other terms $l' \ne -q$ is given by $2\hat P_{2\nu}(0)/[\hat P_{2\nu}(l'\omega+q\omega) + \hat P_{2\nu}(-l'\omega-q\omega)]$.
The recent exploration of PASN in the THz regime \cite{Parmentier16} suggests that our results could be tested in the near future in EQO experiments at fractional filling factor \cite{Glattli16_pss}. 

\section{Conclusions}

We have considered the PASN in a FQH system generated by periodic voltage pulses impinging on a QPC. Due to the typical non-linear behavior of the Green's function in the FQH regime, the photoassisted expression for the noise can be approximated in a remarkably simple way. This allows for a full spectroscopy of the photoassisted probabilities by varying both the dc and the ac amplitude of the voltage pulses. The spectroscopic technique presented in this work is within reach of current experimental technologies.

\section*{References}
\bibliography{levitons_biblio_LT28}

\end{document}